\documentclass{elsarticle}
\usepackage{import}
\usepackage{graphicx}
\usepackage{orcidlink}
\usepackage{soul}
\usepackage{longtable}
\usepackage{geometry}
\usepackage{appendix}
\usepackage{aas_macros}
\usepackage{hyperref}

\begin{document}
\begin{frontmatter}
\title{SDHDF: A new file format for spectral-domain radio astronomy data}

\author[a]{Toomey, L. J.\orcidlink{https://orcid.org/0000-0003-3186-3266}}
\author[a]{Hobbs, G.\orcidlink{https://orcid.org/0000-0003-1502-100X}}
\author[b]{Price, D. C.\orcidlink{https://orcid.org/0000-0003-2783-1608}}
\author[a,c]{Dawson, J. R.\orcidlink{0000-0003-0235-3347}}
\author[h,l]{Wenger, T.}
\author[h]{Lagoy, D.}
\author[d,e]{Staveley-Smith, L.}
\author[f,j]{Green, J. A.}
\author[g]{Carretti, E.}
\author[a]{Hafner, A.}
\author[j]{Huynh, M.}
\author[a,i]{Kaczmarek, J.\orcidlink{0000-0003-4810-7803}}
\author[i]{Mader, S.}
\author[a]{McIntyre, V.}
\author[a]{Reynolds, J.}
\author[h]{Robishaw, T. \orcidlink{0000-0002-4217-5138}}
\author[i]{Sarkissian, J.}
\author[j]{Thompson, A.}
\author[j,k]{Tremblay, C.\orcidlink{0000-0002-4409-3515}}
\author[a]{Zic, A.}

\affiliation[a]{organization={CSIRO Space \& Astronomy},
            addressline={P.O. Box 76}, 
            city={Epping},
            postcode={1710}, 
            state={NSW},
            country={Australia}}
\affiliation[b]{organization={International Centre for Radio Astronomy Research}, 
            addressline={Curtin University},
            city={Bentley},
            postcode={6102},
            state={WA},
            country={Australia}}
\affiliation[c]{organization={School of Mathematical and Physical Sciences and MQ Research Centre in Astronomy, Astrophysics, and Astrotechnology}, 
            addressline={Macquarie University},
            postcode={2109},
            state={NSW},
            country={Australia}}
\affiliation[d]{organization={International Centre for Radio Astronomy Research (ICRAR)}, 
            addressline={University of Western Australia},
            postcode={6009},
            state={WA},
            country={Australia}}
\affiliation[e]{organization={ARC Centre of Excellence for All Sky Astrophysics in 3 Dimensions (ASTRO 3D)},
            country={Australia}}
\affiliation[f]{organization={SKA Observatory}, 
            addressline={SKA-LOW Science Operations Centre, ARRC Building, 26 Dick Perry Avenue, Technology Park},
            city={Kensington},
            postcode={6151},
            state={WA},
            country={Australia}}
\affiliation[g]{organization={INAF, Istituto di Radioastronomia},
            addressline={Via Gobetti 101},
            city={Bologna},
            postcode={40129},
            country={Italy}}
\affiliation[h]{organization={Dominion Radio Astrophysical Observatory}, 
            addressline={Herzberg Astronomy and Astrophysics Research Centre, National Research Council Canada, PO Box 248, Penticton},
            postcode={BC V2A 6J9},
            country={Canada}}
\affiliation[i]{organization={CSIRO Space \& Astronomy},
            addressline={Parkes Observatory}, 
            city={Parkes}, 
            postcode={2870},
            state={NSW},
            country={Australia}}
\affiliation[j]{organization={CSIRO Space \& Astronomy},
            addressline={PO Box 1130}, 
            city={Bentley},
            postcode={6102},
            state={WA},
            country={Australia}}
\affiliation[k]{organization={SETI Institute},
            addressline={Mountain View}, 
            postcode={CA 94043},
            country={USA}}
\affiliation[l]{organization={NSF Astronomy \& Astrophysics Postdoctoral Fellow},
            addressline={Department of Astronomy, University of Wisconsin}, 
            city={Madison},
            postcode={WI 53706},
            country={USA}}

\begin{abstract}
Radio astronomy file formats are now required to store wide frequency bandwidths and multiple simultaneous receiver beams and must be able to account for versatile observing modes and numerous calibration strategies. The need to capture and archive high-time and high frequency-resolution data, along with the comprehensive metadata that fully describe the data, implies that a new data format and new processing software are required. This requirement is suited to a well-defined, hierarchically-structured and flexible file format. In this paper we present the Spectral-Domain Hierarchical Data Format  (`SDHDF') --- a new file format for radio astronomy data, in particular for single dish or beam-formed data streams. Since 2018, SDHDF has been the primary format for data products from the spectral-line and continuum observing modes at Murriyang, the CSIRO Parkes 64-m radio telescope, and we demonstrate that this data format can also be used to store observations of pulsars and fast radio bursts.
\end{abstract}

\begin{keyword}
Radio astronomy
Astronomy data acquisition
\end{keyword}
\end{frontmatter}
 
\section{INTRODUCTION}
\label{sec:intro}
With increasing instantaneous bandwidth and higher data volumes output from modern receivers on radio telescopes, such as the Ultra-Wide-Bandwidth Low Frequency (UWL, \cite{2020PASA...37...12H}) receiver, and the cryogenically-cooled phased-array feed (`Cryo-PAF') receiver at Murriyang\footnote{In the Wiradjuri Dreaming, Biyaami (Baiame) is a prominent creator spirit and is represented in the sky by the stars which also portray the Orion constellation. Murriyang represents the `Skyworld' where Biyaami lives.} the CSIRO Parkes 64-metre radio telescope, there was a requirement for a new file format. This requirement is not unique to Parkes; there are many other observatories with instruments that produce multiple beams and wide bandwidths (such as the Five-hundred-meter Aperture Spherical Telescope `FAST' in China, which utilises similar receiver technology, and has historically written data out with similar formats to Parkes). Here we present Spectral-Domain Hierarchical Data Format (`SDHDF') -- a new file format for radio astronomy data based on the Hierarchical Data Format\footnote{\url{https://www.hdfgroup.org/}} (`HDF\footnote{Note: At the time of writing, any reference to HDF in this paper is to the data model of HDF version 5 (HDF5)}').

Previously, spectral data from Murriyang were stored in `SDFITS' format (\cite{2000ASPC..216..243G}). This format for `Single Dish' data uses the binary table data structure as specified in the Flexible Image Transport System (`FITS', \cite{2010A&A...524A..42P}, \cite{1979ipia.coll..445W}). While the FITS format has enjoyed widespread usage over several decades, design decisions that were reasonable in 1980, now limit modern astronomical research; a summary of these issues is given by \cite{2014ASPC..485..351T, 2015A&C....12..133T}, and we discuss these later on.

The use of HDF as a file format for Parkes spectral-line and continuum data was first introduced specifically for the HI-Pulsar `HIPSR' digital backend (\cite{2016JAI.....541007P}) of the 21-cm multi-beam receiver, as an alternative to SDFITS. However, the HIPSR format was not intended for use outside the project, and does not implement features required for broader use cases.
Amongst the wider astronomy community, HDF has been implemented at the Low Frequency Array (LOFAR, \cite{2013A&A...556A...2V}) for a decade or so (\cite{2011ASPC..442...53A}), albeit more recently for beam-formed data products\footnote{\url{https://support.astron.nl/LOFARBeamformedCookbook/index.html}}. The Laser Interferometer Gravitational-Wave Observatory (LIGO) Scientific Collaboration publish their data in HDF (for example, the O2 Data Release\footnote{\url{https://doi.org/10.7935/CA75-FM95}}), and the format is also supported for image cube visualisation of large-scale data from several telescopes such as the Karl G. Jansky Very Large Array (VLA), the Atacama Large Millimeter/sub-millimeter Array (ALMA), and SKA pathfinders, with the Cube Analysis and Rendering Tool for Astronomy (CARTA\footnote{\url{https://doi.org/10.5281/zenodo.3377984}}) (\cite{2020A&C....3200389C}).

Consequently we explored the use of HDF for all spectral-line and continuum data products from Parkes receivers, converging on the SDHDF model as the formal convention based on the following considerations:

\begin{itemize}
\item The storage mechanism is based on HDF, a well-defined and documented format. The HDF Group states that they are ``committed to always keeping HDF software free and open source'', and this is reflected in the extensive and up-to-date commit history of the repository.
\item SDHDF is capable of storing comprehensive high time-resolution and wide-band high frequency-resolution data in a nested hierarchical structure suitable for telescope feeds with one beam (single pixel feed) or multiple beams, and with the potential to expand to support beam-formed data products from multiple single dishes, antennas or tiles.
\item SDHDF data can be incorporated into HDF Virtual Datasets (VDS\footnote{\url{https://docs.h5py.org/en/stable/vds.html}}), and the format is suitable for working with terabyte-scale data products in distributed computing infrastructure.
\item The data and metadata are fully self-described by HDF `attributes'.
\item The format is designed to be portable and open to new functionality and capability, with long-term development support funded by donation from within the community.
\item Reduction of the data products can produce secondary data products that themselves meet the SDHDF definition.
\item The format can be parsed by HDF libraries available for computer languages such as C/C++, MATLAB, IDL, Julia, Python, as well as by open source tools and graphical user interfaces (GUIs).
\item SDHDF data products are suitable for storage in long-term archives because the metadata are described, the files are compressible, and the SDHDF definition has been allocated a Digital Object Identifier\footnote{\url{https://www.doi.org}} (DOI) for the life-cycle of the data.
\end{itemize}

SDHDF was successfully commissioned as the primary format for all spectral-line and continuum data products on completion of the Parkes UWL receiver project. High time-resolution pulsar observations are still produced in `PSRFITS' format (\cite{hotan_van_straten_manchester_2004}), but we demonstrate below that including pulsar data in SDHDF format would be possible in the future. Since late 2018, more than 80 TB of SDHDF data products have been archived in the Australia Telescope Online Archive\footnote{\url{https://atoa.atnf.csiro.au/}} (ATOA), becoming publicly available after an initial 18 month embargo period.

In this paper, we provide an overview of the SDHDF definition in Section \ref{sec:sec2}, and present design features in Section \ref{sec:sec3}. In Section \ref{sec:sec4} we provide information on working with SDHDF files, including the introduction of a new software package `INtegrated SDHDF Processing Engine in C for Telescope data Analysis'  (\textsc{INSPECTA}), developed specifically for working with this format and for which we provide some example use cases. In Section \ref{sec:sec5} we discuss the choice of HDF as the base for our format, further SDHDF development, and uptake within the community. We present our conclusion in Section \ref{sec:concl}. An overview of the definition (Appendix A), and an example of how the format stores some observation parameters for Parkes data (Appendix B), are available in the online supplementary material.

\section{THE SDHDF DEFINITION}
\label{sec:sec2}
\begin{figure*}
\centering
\includegraphics[width=450px]{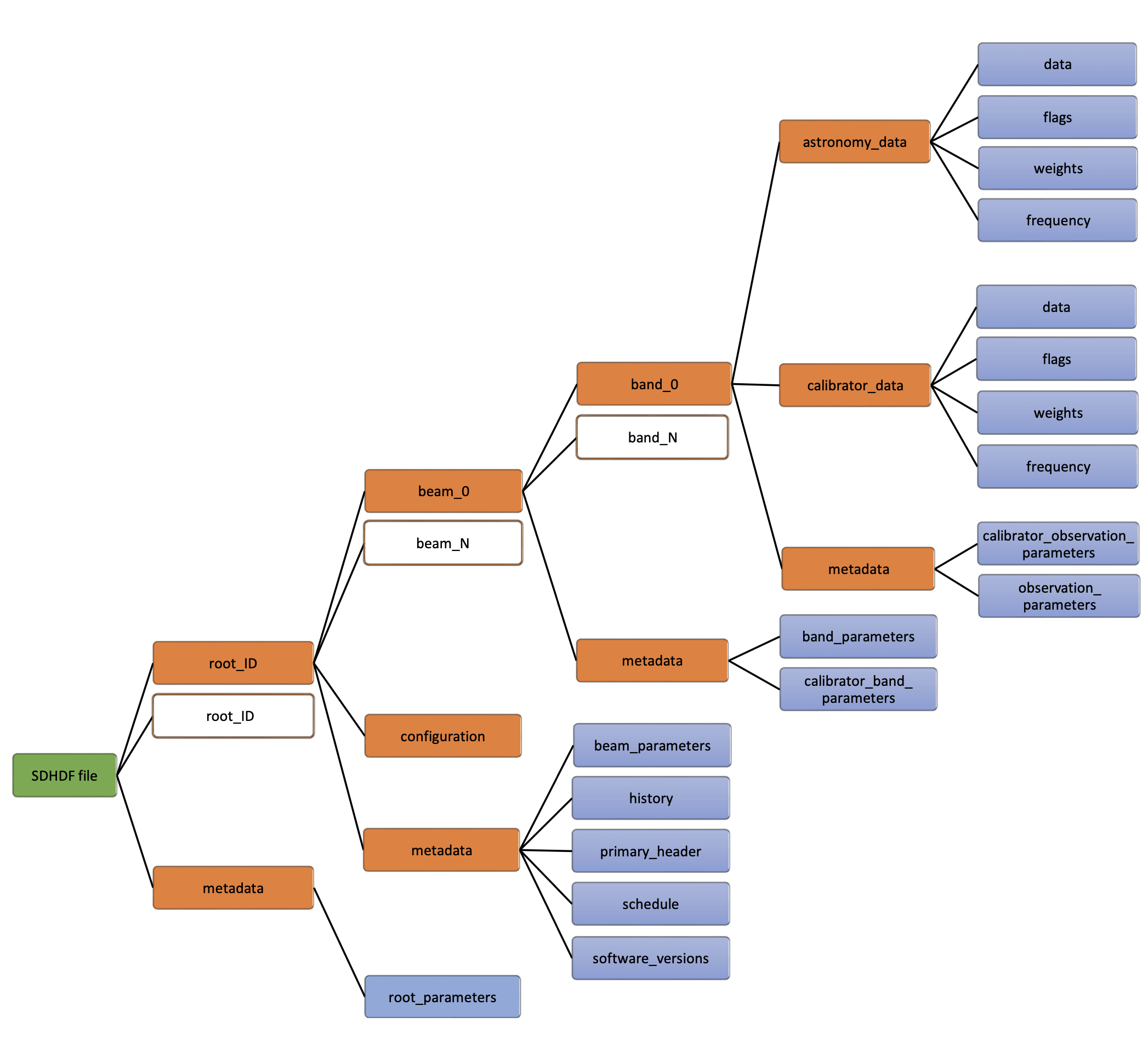}
\caption{A visual representation of an example data file, showing the hierarchical association of HDF groups and datasets. An SDHDF data file (the solid green rectangle) may contain optionally none or any number of `root\_ID' objects each of which is of the SDHDF\_CLASS {\tt{sdhdf\_file}} and each adhering to the SDHDF definition, where each root\_ID object could be a separate observation and grouped together (for example one may wish to store multiple single scans in the same file). Information about the root\_ID objects are stored in a `root\_parameters' dataset; this dataset and the parent metadata group are not required if the file does not contain a root\_ID object. In this Figure, the solid orange and solid blue rectangles represent HDF group and dataset objects respectively. Any number of root\_ID, beam and band groups per beam are supported, as illustrated by the unfilled rectangles with orange borders. For Parkes UWL data, we chose to record a switching noise source for calibration purposes, storing both on, off and a 32-bin time-averaged dataset in the `calibrator\_data' group (not shown in this Figure).}
\label{fg:sdhdf_structure}
\end{figure*}

The SDHDF definition can be thought of as a tree-like structure of HDF `group' and `dataset' binary objects, containing the data, observation metadata, and time, frequency and polarisation information. A visual representation of the structure is shown in Figure \ref{fg:sdhdf_structure}.

SDHDF was designed for flexibility and to accommodate multi-dimensional channelised data in one or many frequency bands of any bandwidth in blocks of time of sub-second precision. A single file can contain one or many observations from different observing modes (for example a targeted pointing and/or scan) stored as multiple top-level `root\_ID' objects, each of which adhere to the SDHDF definition and can be a combination of many input data streams from one or many beams, associated metadata for both astronomy and noise source data products, and observation metadata. An SDHDF file may also be a product of post-processing from any number of other SDHDF data and metadata products structured according to the definition -- these may contain more structure than the raw data. For example, we may optionally store datasets that have had mean or median averaging applied, or store additional zoomed frequency bands. The SDHDF definition\footnote{https://doi.org/10.25919/880t-0m95} and a template data file are publicly available on CSIRO's Data Access Portal\footnote{\url{https://data.csiro.au/}}.

In order to ensure that all SDHDF data products from a set of telescope observations meet the requirements of the definition, we have implemented a template methodology. With this,
on completion of an observation, the data and metadata structure, descriptions, data units and values are first populated in an SDHDF template file built from the definition, where parameters are configured for a specific telescope and observing system. The HDF datasets for the astronomy data are initiated such that they are resizable. The template structure at this stage is just a placeholder -- it is then populated by the raw data and metadata to form the final data product. Metadata are stored with HDF `attributes' -- objects that directly describe the dataset or group object to which they are attributed. All attributes are stored as native Python data types.

The benefits of a flexible and configurable template structure are two-fold --- it facilitates uptake of the format by a range of radio astronomy institutions, and use of a template ensures that all data products from different institutions adheres to the SDHDF format definition, thereby enhancing data provenance in long-term archives.

\subsection{SDHDF classes and attributes}
The SDHDF definition contains HDF classes and attributes that are unique to an SDHDF format file -- these are useful for grouping objects together when performing operations on the data.
Classes are assigned to specific dataset types, identified by the `SDHDF\_CLASS' attribute. Table \ref{tb:sdhdf_classes} lists some examples of these classes and their descriptions.

Two classes, {\tt{sdhdf\_table}} and {\tt{sdhdf\_waterfall}} are central to the SDHDF definition. We provide a brief summary below; further details about these and other SDHDF classes can be found in the SDHDF definition. Observation metadata are stored in tables with the SDHDF\_CLASS {\tt{sdhdf\_table}}. This class is similar to the HDF Table Specification Version 1.0\footnote{\url{https://docs.hdfgroup.org/hdf5/develop/_t_b_l.html}}.  A one-dimensional compound HDF dataset is used to store the data,  with column names ascribed in the compound datatype.  
 
The {\tt{sdhdf\_waterfall}} class is an $N$-dimensional array with dimension scales to describe each axis.  Following the HDF dimension scale specification, the attribute `DIMENSION\_LABELS' provides a list of strings to describe each axis. Similarly, the attribute `DIMENSION\_LIST' provides references to any datasets that represent a dimension scale.  While the {\tt{sdhdf\_waterfall}} class can be used for any $N$-dimensional dataset, in this context the dimensions are labelled as time, product type (default: polarisation), frequency and bin\footnote{In the {\tt{sdhdf\_waterfall}} class, the term `bin' refers to the phase of a periodic signal, such as that of a pulsar or switched noise source}, as defined in Section~\S\ref{sec:flexibleObserving}. Information about units (e.g. jansky, kelvin, counts) is stored in the `UNIT' attribute. 

Arrays in the {\tt{sdhdf\_waterfall}} datasets can be large and HDF provides the means to store multiple segments separately in the file (using a method known as `chunking').  The SDHDF definition does not preclude chunking any dataset in a way that will make reading and writing the data more efficient.  For the Parkes observations we currently use an automatic chunking algorithm, but are currently exploring more efficient methods.   We note that as long as HDF tools are used then the choice of chunking method does not affect how the data sets are loaded or processed.

Throughout the definition, other attributes are used to define special parameters --- for example, the `FRAME' attribute describes the spectral reference frame of the frequency dataset, set to `topocentric' by default in the definition. 

\begin{table}
  \caption{Examples of SDHDF classes}
  \begin{center}
  \begin{tabular}{ll}
    Class name & Description \\
\hline
\\
{\tt{sdhdf\_file}} & SDHDF file class \\
{\tt{sdhdf\_band}} & SDHDF frequency band class \\
{\tt{sdhdf\_table}} & SDHDF metadata table class \\
{\tt{sdhdf\_frequency}} & SDHDF frequency dataset \\
{\tt{sdhdf\_data}} & SDHDF astronomy dataset \\
{\tt{sdhdf\_waterfall}} & SDHDF frequency time dataset \\
{\tt{sdhdf\_flags}} & SDHDF RFI flags dataset \\
{\tt{sdhdf\_weights}} & SDHDF RFI weights dataset \\
\hline
\\
  \end{tabular}
  \end{center}
  \label{tb:sdhdf_classes}
\end{table}

\begin{table}
  \caption{Example metadata from an SDHDF dataset object}
  \begin{center}
  \begin{tabular}{p{1.6in}p{2in}p{0.8in}p{1.3in}}
    Path Information & & & \\
\hline
\\
Basename & beam\_parameters & & \\
Local & sdhdf\_template.hdf::\//metadata\//beam\_parameters & & \\
Physical & sdhdf\_template.hdf::\//metadata\//beam\_parameters & & \\
\\
Data Information & & & \\
\hline
\\
HDF type & COMPOUND & & \\
\\
Data types & LABEL (string) & & \\
& NUMBER\_OF\_BANDS (int64) & & \\
& SOURCE (string) & & \\
& RIGHT\_ASCENSION (string) & & \\
& DECLINATION (string) & & \\
\\
Default Attributes & & &\\
\hline
Name & Description & Unit & Value \\
\hline
\\
SDHDF\_CLASS & SDHDF class name & None & sdhdf\_table \\
SDHDF\_DESCRIPTION & SDHDF object description & None & Metadata specific to the antenna beam \\
\\
Dimension Attributes & & &\\
\hline
Name & Description & Unit & Value \\
\hline
\\
LABEL & Beam label & None & beam\_N \\
NUMBER\_OF\_BANDS & Number of frequency bands & None & band\_N \\
SOURCE & Source name & None & source\_name \\
RIGHT\_ASCENSION & Coordinates in Right Ascension & HH:MM:SS.s & HH:MM:SS.s \\
DECLINATION & Coordinates in Declination & DD:MM:SS.s & DD:MM:SS.s \\
\\
Additional Attributes & & &\\
\hline
Name & Description & Unit & Value \\
\hline
\\
EQUINOX & Equinox of the coordinates & Julian year & NNNN\\
\hline
  \end{tabular}
  \end{center}
  \label{tb:metadata}
\end{table}

In Table \ref{tb:metadata} we show an example of the metadata including attributes, associated with a dataset that contains information about a telescope beam.  The position of the dataset in the object hierarchy is shown in `Path Information', with the data being stored as a compound entity of two different data types (`Data Information'). Attributes of the example dataset include `NUMBER\_OF\_BANDS' and `SOURCE' entries -- the definition allows for any number of beams, each with potentially different sources and associated metadata for any number of frequency bands (sub- or zoom-bands).

\section{SDHDF DESIGN FEATURES}
\label{sec:sec3}
\subsection{Flexible observing modes}\label{sec:flexibleObserving}
The SDHDF structure is flexible enough to accommodate various observing modes. The telescope may scan across a sky region, the receiver may rotate to account for parallactic angle, or (with receivers such as phased-array feeds) individual beams may be steered independently. The top (or `root\_ID' level) in the SDHDF definition stores a description of each receiver beam (0 to $N$-1 beams, where `beam\_0' is the first beam\footnote{In SDHDF, all indices are zero-based.}).

The tier down from the beam group is the `band' group object.  In this context a `band' refers to a single range of frequencies, and may be colloquially referred to as a sub- or zoom-band. Note that we do not explicitly preclude overlapping bands; for instance a file could include a wide bandwidth and a separate band that is a sub-set of the full band. The information stored in the band group will have originally come from a signal processor, which will have formed spectra at specific time intervals.

We store time-dependent information (such as the timestamp of each spectrum, pointing information, weather conditions, etc.)  independently for each telescope beam and observing band in the `observation\_parameters' dataset. For Parkes data we store this information with a default cadence corresponding to the rate that the time-averaged spectra are recorded, but the implementation is flexible enough to store information on any cadence. The `observation\_parameters' dataset is configurable based on the requirements of a particular telescope -- an example of the parameters required for a Parkes observation is shown in \ref{appendix:appdxB}. Metadata information, such as the time corresponding to a specific spectrum, or the frequency of a given frequency channel, is set to the centre of the time interval or the frequency channel. The primary time information is stored as an integer Modified Julian Date (MJD) and the number of seconds (as a float) since the start of that MJD. Storing these as separate values allows the format to have sufficient time precision for pulsar data.

The spectra are stored for each beam and band in the `astronomy\_data' groups, as a 4-dimensional {\tt{sdhdf\_waterfall}} dataset with floating point precision. For the majority of continuum and spectral-line observations only a single phase bin is stored, however, for pulsar or calibration observations a large number of phase bins may be recorded.  The frequency of each channel is stored in the frequency dataset.  This is a two dimensional array with the dimensions being the number of integrations and the number of frequency channels.  In many cases only a single integration is present (as the frequency of the channels do not change in time). However, if the frequency axis has been Doppler corrected to account for the Earth's motion then the number of integrations likely will equal the number of spectra in the file.

The spectra recorded in the SDHDF file usually need to be calibrated.  The calibration steps depend on the observing system and the particular science case. The SDHDF structure therefore needs to be sufficiently flexible to allow for different calibration methods.  SDHDF was initially developed for the spectral-line and continuum output data from the Parkes telescope, in which the astronomy spectrum is recorded with high time and frequency resolution, and spectra with the noise source switched on and separately switched off are also recorded with lower time and frequency resolution.    Such information is stored in datasets for each beam and band in the `calibrator\_data' group with {\tt{sdhdf\_waterfall}} classes being used to store the on and off spectra for Parkes data.  In order to confirm that the signal processor was correctly synchronised with the noise source we also used the {\tt{sdhdf\_waterfall}} class with multiple phase bins to record spectra as measured in 32 phase bins across the noise source period. Details of the calibration data can also be stored as HDF dataset attributes. For instance, the switched noise source data may be normalised so that the counts in the noise source spectra are equivalent to counts in the astronomy spectrum.

Traditional spectral-line calibration strategies rely on switching between observation parameters. For instance, the telescope position may switch between an on-source and an off-source position (or swapping beam positions), or the signal-processing system may implement a frequency-switching process. The current Parkes signal processor produces separate SDHDF data files for different source positions (i.e., on- and off-source pointings are in different files), however as shown in Figure~\ref{fg:sdhdf_structure}, an SDHDF file can have multiple `root\_ID' groups. These can be used to store multiple observations (or multiple observing setups) in a single SDHDF file.  Frequency-switched observations typically shift the frequency of a local oscillator in order to facilitate spectral bandpass calibration. Within SDHDF, such observations could be stored as (1) individual SDHDF files for each change in frequency, (2) as separate `root\_ID' groups for each frequency, (3) by adding in an extra `scan' hierarchy after the `band' group, (4) by expanding the `observation\_parameters' metadata to include the local oscillator frequency or (5) with the inclusion of observatory-specific calibration datasets to be included in the `calibrator\_data' group. See  Section~\ref{sec:extension} on extending the format.

\subsection{Additional datasets, flags and weights}
The SDHDF definition includes scope for extra, optional datasets. For instance, these could be required as part of the data reduction process.  Perhaps the most commonly used extra datasets relate to weighting and flagging the spectra.  The output of manual, or automatic, flagging routines can be stored in the `flags' datasets. This is recorded with a 0 to indicate a spectral channel is not flagged and a 1 otherwise. Unless specified explicitly we use unsigned 8-bit values to record these flags to enable simple reading and writing, but they can be written as 1-bit values if required. We allow for flags to be applied to astronomy or calibration spectral channels or integrations. The PSRFITS definition does not allow flags to be applied independently for polarisation channels. However, we provide the option that flags (and weights, as described below) can be written with the same dimensions as the raw data (including polarisation and bin information). However if, for instance, the polarisation dimension is set to 1 then the flags would be applied to all polarisations for a given time and frequency channel.

Weights can also be applied (usually to each frequency channel in each integration).  The weight can represent the product of the channel bandwidth and integration time (or other weighting methods could be used such as \cite{anderson2023methods}).  SDHDF weights are stored as floating point values in the `weights' datasets.

If the data are recorded with low-bit quantisation (for example, pulsar search-mode data is generally a 2-bit data stream), then the data type is stored in the `DATA\_TYPE' attribute of the {\tt{sdhdf\_waterfall}} class. It is often necessary to provide tables of scaling factors enabling the conversion of the stored bit depth back to the measured signal strength.  Following the PSRFITS definition we allow for independent {\tt{sdhdf\_waterfall}} datasets containing offsets and scaling parameters where the measured value $v$ is recovered from the stored unsigned integer data value $d$ as:
\begin{equation}
    v = (d - z)S + \zeta
\end{equation}
where $z$ is a `zero-offset' value, $S$ a frequency, polarisation and time-dependent scaling factor and $\zeta$ a frequency, polarisation and time-dependent offset.

\section{WORKING WITH SDHDF DATA}
\label{sec:sec4}
Open source GUIs and Python modules are available for interrogating HDF files in general, and the Parkes spectral-line and continuum teams have also been independently developing their own SDHDF processing tools in Python. We have developed the comprehensive \textsc{INSPECTA} software package, included as part of the publicly available \textsc{SDHDF\_TOOLS}\footnote{\url{https://bitbucket.csiro.au/scm/cpda/sdhdf_tools}} repository. \textsc{INSPECTA} is written in C and contains tools to perform a multitude of processing tasks, from visual inspection of files to calibration and even the simulation of various observing parameters. The complete list of tools is too comprehensive to list here, but allows for spectra to be visualised, calibrated and analysed.  It also allows for both automatic and manual radio-frequency interference (RFI) flagging. We provide examples below in which SDHDF files have been processed using the \textsc{INSPECTA} software to demonstrate the suitability of SDHDF for different types of observations.  These observations have been carried out with Murriyang and the majority make use of the UWL receiver.

\begin{figure}
    \centering
    \includegraphics[width=7.5cm]{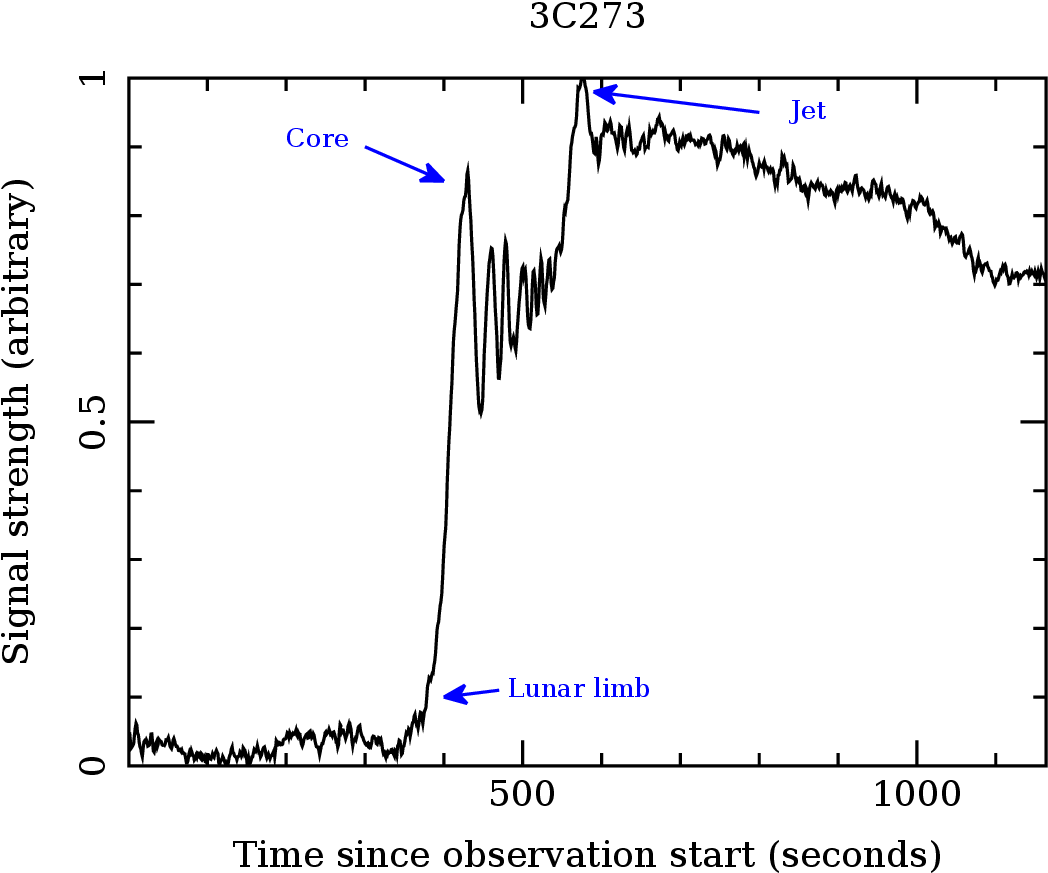}
    \includegraphics[width=7.5cm]{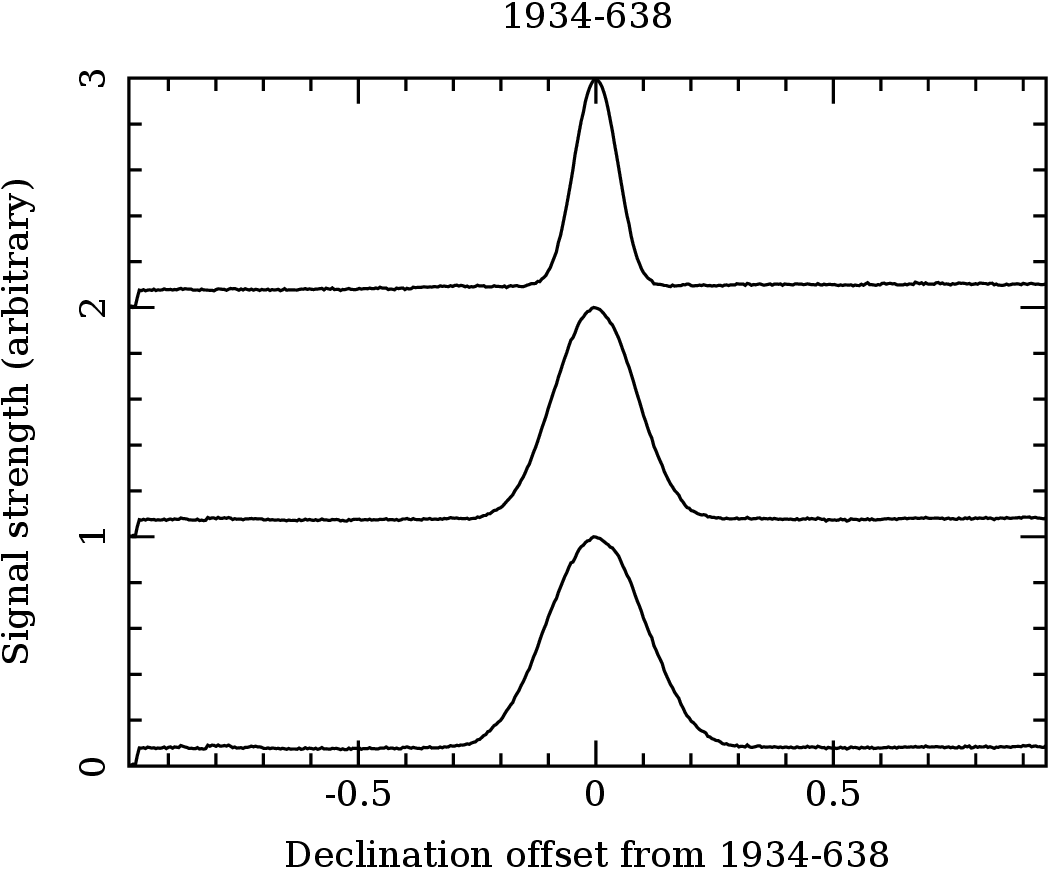}
    \caption{(Left panel) The signal strength in the UWL 1152 MHz band as a function of time during the lunar occultation of the quasar 3C 273. The quasar core and then a jet boost the flux as they emerge from behind the lunar limb. (Right panel) Representation of a scan across a primary flux calibrator source PKS~1934$-$638 in three observing bands centred on 3328 (top), 1664 (centre) and 1408\,MHz (bottom), where the declination offset is in degrees.}
    \label{fig:timeDependence}
\end{figure}

\begin{figure}
    \centering
    \includegraphics[width=16cm]{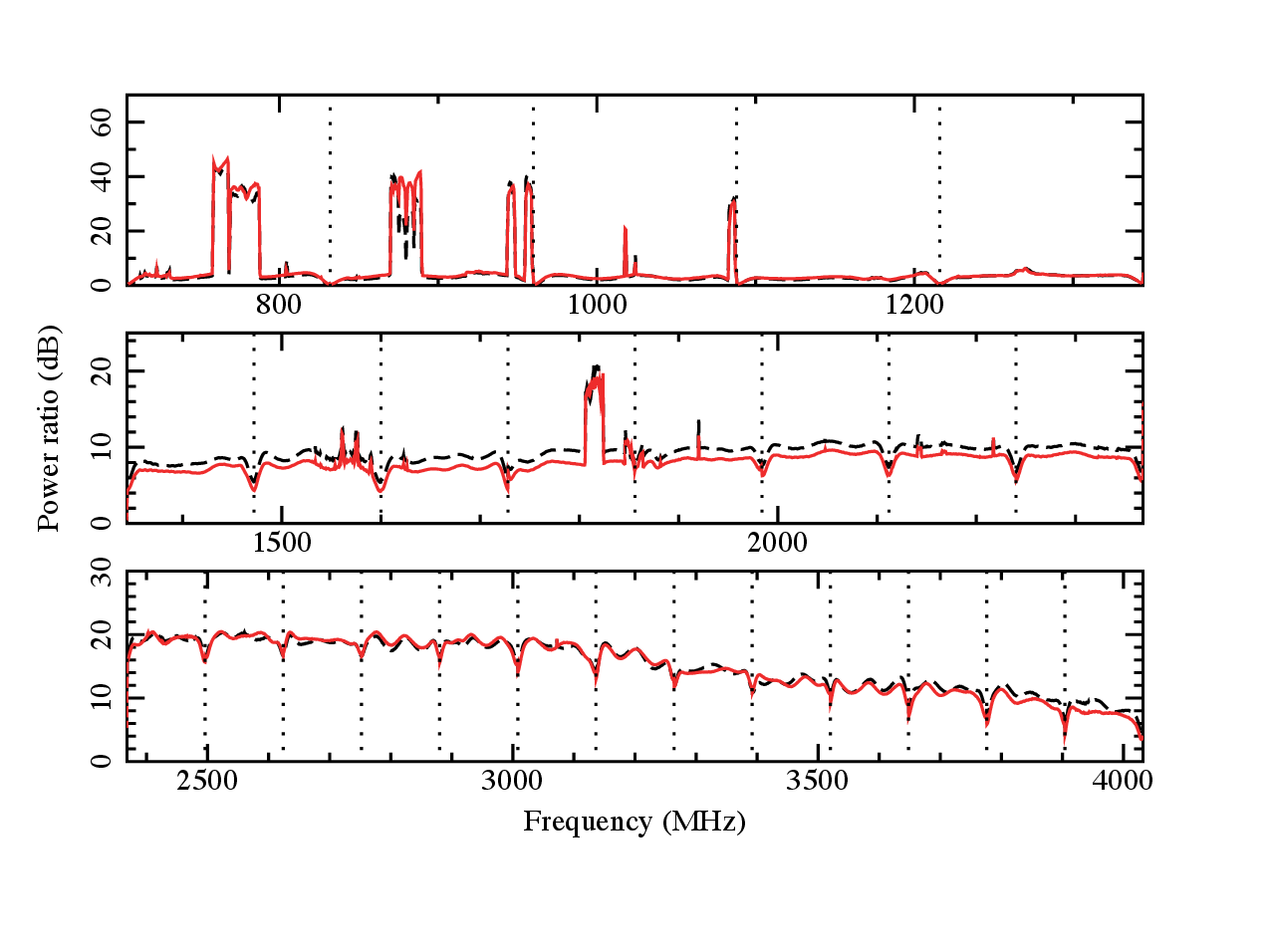}
    \caption{A wide-bandwidth observation demonstrating the ability of SDHDF to store, process and visualise such data. The three sub-panels represent the three RF bands recorded by the Parkes UWL receiver system (704 to 1344 MHz, 1344 to 2468 MHz and 2468 to 4032 MHz), showing polarisations A and B in red and black (dashed) respectively.}
    \label{fig:wideband}
\end{figure}

Here we show both the versatility of the SDHDF format and the \textsc{INSPECTA} software package by demonstrating how a diverse set of observations can be processed and visualised. An example is shown in the left panel of Figure~\ref{fig:timeDependence} in which the telescope was tracking the limb of the Moon as it occulted quasar 3C 273 on 20th February 2022 (as a modern re-enactment of \cite{1963Natur.197.1037H}). The SDHDF file for this observation was obtained from the ATOA\footnote{\url{https://atoa.atnf.csiro.au/}; Project ID: PX083; File: uwl\_220219\_143803\_3.hdf}. We recorded observations with 400\,Hz resolution producing over 7 million spectral channels over the entire UWL band, which were written to disk each second. In order to ensure we do not produce a single file that is too large to be archived, our online processing scripts split data files when they reach a specified size (currently $\sim 10$\,GB) -- the files can be split by beam, band or time, and the \textsc{INSPECTA} software can subsequently be used to join files back together as required.   The total observation time for this specific data file was $\sim 1160$ seconds. The SDHDF file captures each of the 26 sub-bands formed by the UWL receiver system along with metadata recording the telescope pointing directions for each of the 1\,s integrations.  To form the output shown in Figure~\ref{fig:wideband} we used the \textsc{INSPECTA} tools to average the same observation in time. This is then visualised as the three RF bands for both polarisation channels.  Note that no calibration has been carried out in this stage and hence the Figure shows the frequency-dependent bandpass shape and strong RFI signals primarily from mobile communication systems.

This observation can also be used to demonstrate continuum observations in which a low frequency resolution is generally required, but the data are output with short time duration, showing that the SDHDF format can store pointing information at sufficient precision and temporal resolution. We first carried out an automatic removal of both persistent and transient RFI using the \textsc{INSPECTA} software.   We then averaged the frequency channels, summed the polarisations and plotted the total signal strength as a function of time in the 128 MHz-wide sub-band centred at 1408 MHz. Of course, we can make similar plots in other sub-bands, and a full analysis of the recorded signal will be presented elsewhere.

The occultation observation was carried out by tracking the position of the source; it is more common to scan the telescope beam across a known bright source.  In the right panel of Figure~\ref{fig:timeDependence} we show a scan across the flux density calibrator source PKS~1934$-$638. Here we have written three frequency bands to the SDHDF file (centred on 1408, 1664 and 3328\,MHz) and the traces have been normalised and offset from each other to show the beam widths. This demonstrates that the stored metadata provide sufficient time and positional information to recover the source position. The SDHDF file for this observation was obtained from the ATOA\footnote{\url{https://atoa.atnf.csiro.au/}; Project ID: P974; File: uwl\_210914\_053215.hdf}.

To demonstrate that SDHDF records sufficient information to calibrate the flux density and polarisation properties of a spectral-line source we use a known, bright hydroxyl maser associated with high-mass star formation, G330.953$-$0.182 (\cite{2014MNRAS.439.1680C}, references therein). The SDHDF file was also obtained from the ATOA\footnote{\url{https://atoa.atnf.csiro.au/}; Project ID: P1073; File: uwl\_220323\_193627.hdf}.  For this example we do not use an off-source pointing (although the archive does contain off-source pointings obtained for this observing project).  During the observation the noise source was switched on and off at a frequency of 100 Hz.  The SDHDF file contains the astronomical spectrum as originally recorded, along with the spectra at lower frequency resolution when the noise source is on and also when it is off, separated out using the known phase of the noise source. We note that the integration time of the astronomy data can be different from that of the noise source. We use \textsc{INSPECTA} to average the data in time and to extract the part of the observing band that contains the hydroxyl lines of interest (the 1665.402 and 1667.359 MHz main-line ground-state transitions).  We then use the switching noise source to calibrate the astronomy spectrum (both flux density and polarisation).  This makes use of knowledge of the noise source amplitude measured through separate on- and off-source observations of PKS 1934$-$638. These flux calibration solutions for Murriyang are available online\footnote{\url{https://www.parkes.atnf.csiro.au/observing/Calibration_and_Data_Processing_Files.html}}.

\begin{figure}
    \centering
    \includegraphics[width=13cm]{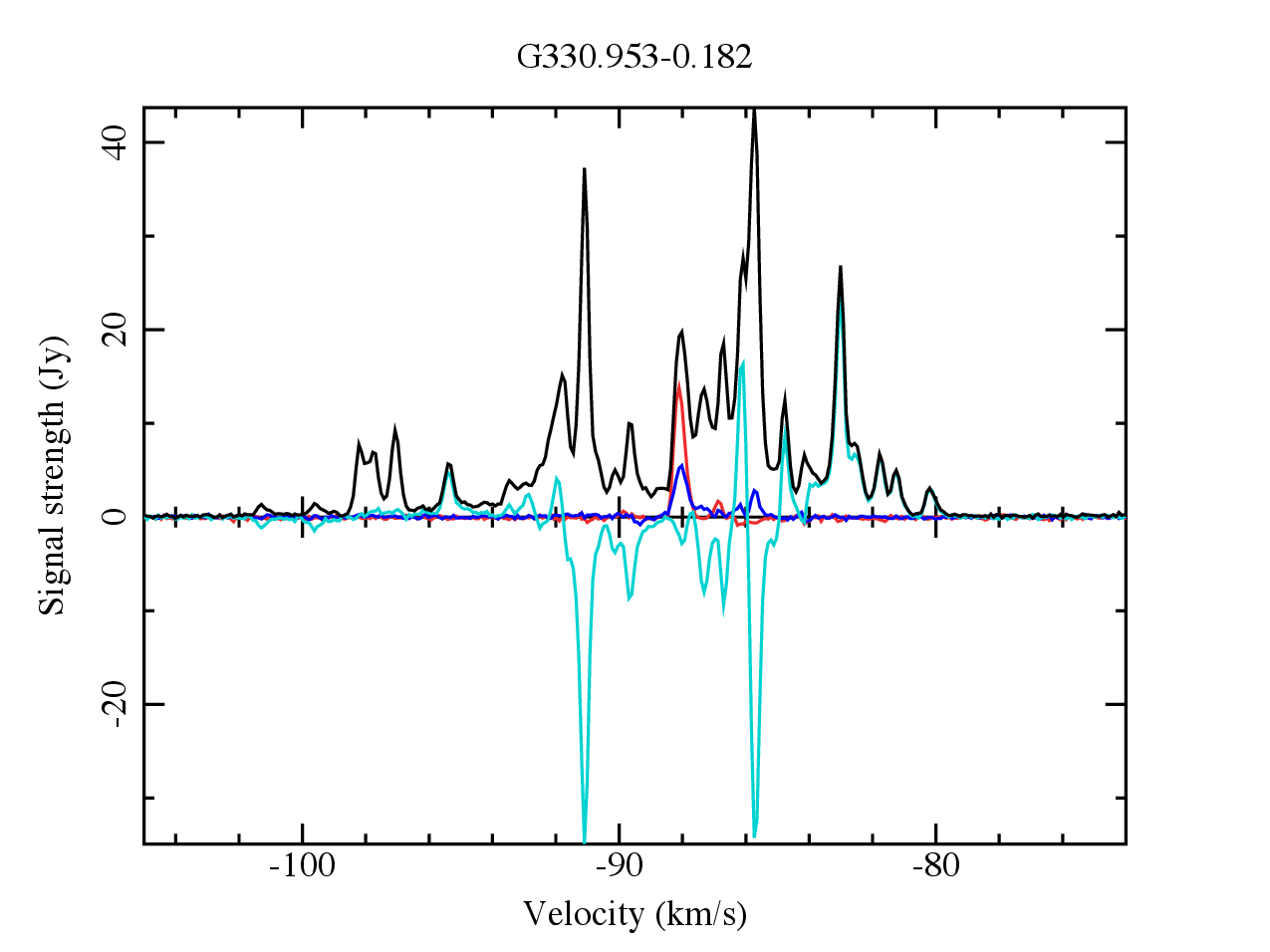} \\
    ~\hspace{-1.2cm}
    \includegraphics[width=11cm]{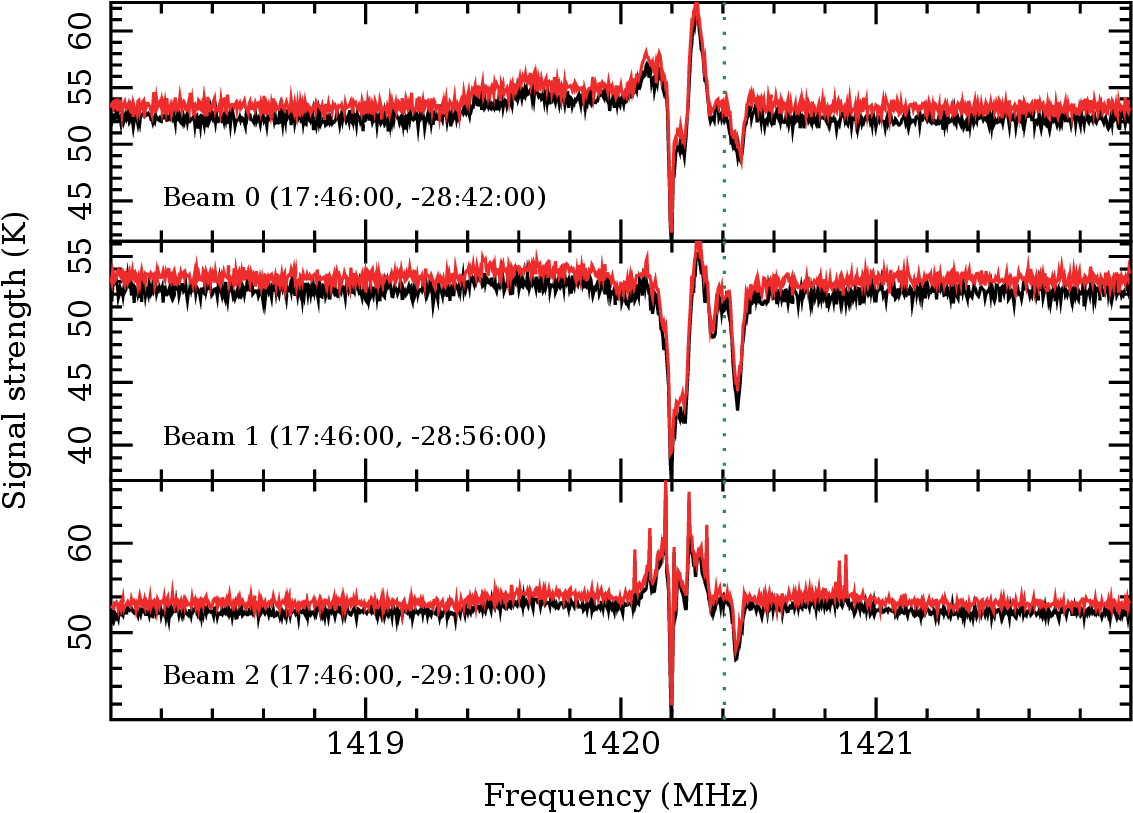}
    \caption{(Top panel) Fully calibrated spectrum of the known 1665-MHz hydroxyl maser G330.953$-$0.182 from a Parkes UWL observation, showing the Stokes parameters I, Q, U and V in black, red, blue and teal respectively. (Bottom panel) Simulated example multiple-beam observations of Galactic hydrogen making use of Parkes Galactic All-Sky Survey (GASS) observations of adjacent positions separated by 14 arc-minutes (the beam full width half maximum) in declination, showing polarisations A and B in red and black respectively.  Note that all beams are stored within the same SDHDF file, which can account for both fixed and moving beam directions.}
    \label{fig:spectralLines}
\end{figure}

The left panel in Figure~\ref{fig:spectralLines} shows the flux and polarisation calibrated spectrum, using the time information stored in the SDHDF file to convert the observed topocentric frequencies to the local-standard-of-rest velocity.   A detailed monitoring study of the variability of this source is being carried out using similar analysis methods and will be published elsewhere.

SDHDF has been designed to allow for multiple antenna beams. However, our test case with the Parkes UWL system only provides a single beam.  In order to demonstrate that archival multi-beam data can be stored in SDHDF format, and that the format is adequate for the future (such as for phased-array feeds) we have used \textsc{INSPECTA} to simulate data of Galactic hydrogen emission and absorption in the vicinity of the Galactic Centre based on input spectra from the Parkes Galactic All-Sky Survey (GASS; \cite{mpc+09}) -- refer to the right panel of Figure~\ref{fig:spectralLines}. We have also demonstrated that the SDHDF format and the \textsc{INSPECTA} software can be used for storing many more beams (we have successfully simulated, stored and processed 72 beams in a single SDHDF file, and the scaling is linear).

\subsection{Storing phase-resolved, high time-resolution data and data precision}
SDHDF was originally developed for the spectral-line and continuum radio astronomy communities.  However, SDHDF can store phase-resolved data and therefore can also be used for pulsar-style observations. Additionally SDHDF has the potential to store multiple modes of observations, such as those from commensal surveys.

For nearly two decades, PSRFITS has been the standard data format for phase-resolved and time-series data such as those from pulsar fold- and search-mode observations respectively. The format is well supported by pulsar processing software packages that are trusted by the community. However, we believe that pulsar data would also benefit from being stored in SDHDF for the following reasons (noting that this is not an exhaustive list):

\begin{itemize}
\item PSRFITS format is not hierarchically structured, and so has no concept of the beam dimension and therefore cannot support multi-beam data.
\item I/O is significantly affected if a PSRFITS binary data unit is appended to, unlike the chunked storage in HDF which allows the file to be easily extended. 
\item For pulsar search-mode observations, there is generally a trade-off between splitting the time-series into small chunks each with its own metadata header (as in the PSRFITS format) and having a single header and an unbroken time series (as in SIGPROC's filterbank\footnote{\url{https://sigproc.sourceforge.net}} format).  The SDHDF format allows for an unbroken time-series, whilst storing the metadata at the required cadence.
\item Pulsar fold-mode observations generally start and/or end with separate calibration observations of the injected noise source, that are currently stored in separate files --- these could all be incorporated into the same SDHDF file for simplicity and ease of use.
\item SDHDF is able to store auxiliary data products such as images of integrated pulse profiles or gain variations which would add value to the data once archived. 
\item The spectral-line and pulsar software packages are independent and yet apply similar methods, for example, for calibration.  A single set of algorithms can be used to ensure consistency between different observing modes.
\end{itemize}

\begin{figure}
    \centering
    \includegraphics[width=10cm]{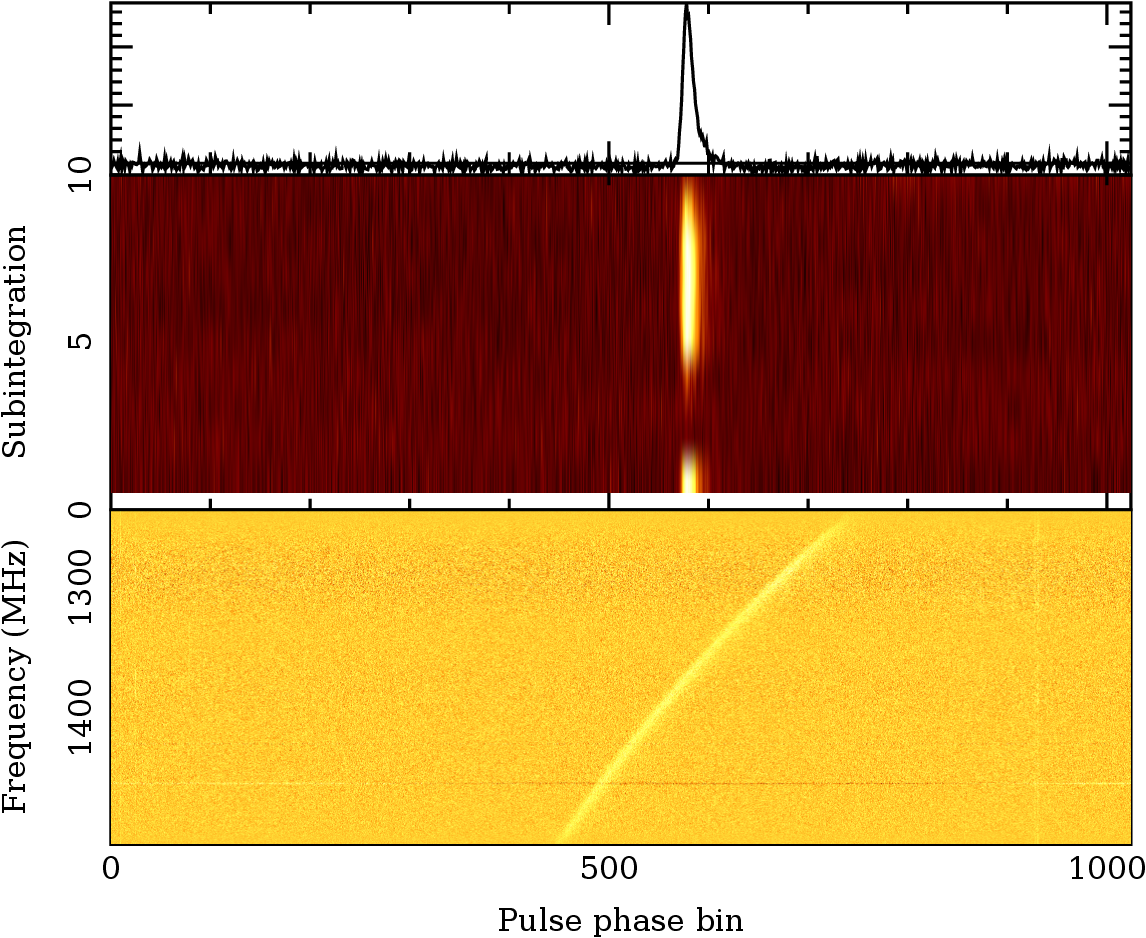}
    \caption{Visualisation of data requiring phase bins. This observation is of the intermittent pulsar PSR~J1717$-$4054 and the data stream has been channelised and folded online at the known period of the pulsar. The top panel shows the de-dispersed and averaged pulse profile. The central panel represents the de-dispersed pulse signal strength as a function of sub-integration, and the bottom panel shows the dispersed pulse signal strength as a function of phase and frequency.}
    \label{fig:foldPulsar}
\end{figure}

Pulsar fold-mode data are produced by folding the incoming signal at the known period of the pulsar and storing the pulse profile for each frequency and polarisation channel with a specified number of phase bins.  Such data can be stored within the SDHDF format without modification as the data structure allows for binned data. An example pulsar profile is shown in Figure~\ref{fig:foldPulsar}. The pulsar, PSR~J1717$-$4054 was observed as part of the PULSE@Parkes outreach project (\cite{2009PASA...26..468H}) --- the PSRFITS file can be obtained from the DAP\footnote{\url{https://data.csiro.au/}; Project ID: P595; File: s101109\_000655.rf}. The file was converted into SDHDF using the \textsc{INSPECTA} tools and then visualised. The upper panel shows the time, polarisation and frequency averaged de-dispersed pulse profile, the central panel shows the pulse amplitude as a function of time (noting that the pulsar switches off and on during the observation) and the bottom panel shows the dispersed pulse sweep as a function of observing frequency.

\begin{figure}
    \centering
    \includegraphics[width=10cm]{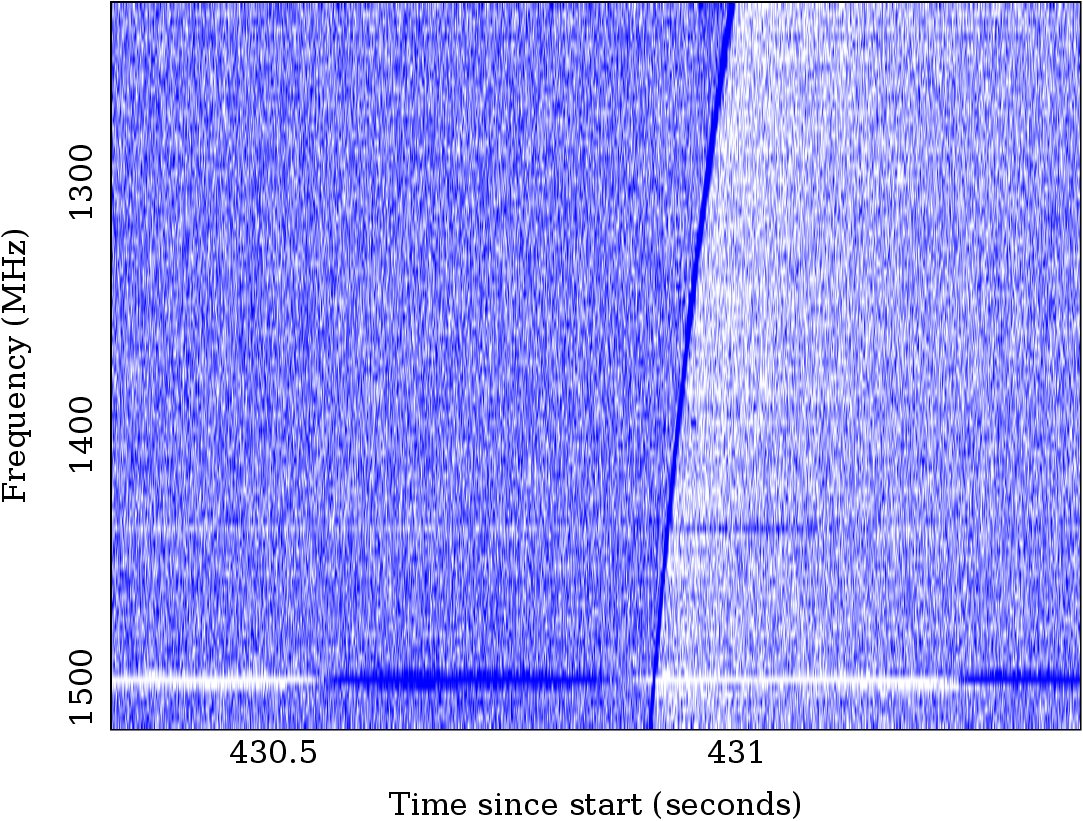}
    \caption{High time-resolution, multi-beam data converted into SDHDF format. We show 1\,second of data around the first discovered fast radio burst (FRB) in beam 6 of the Parkes multi-beam receiver (\cite{doi:10.1126/science.1147532}).}
    \label{fig:searchPulsar}
\end{figure}

High time-resolution time-series data (used, for instance, to search for pulsars or fast radio bursts) cannot be stored with floating point precision.  Typically such data streams are recorded with 1, 2, 4 or 8-bits of precision.  In Figure~\ref{fig:searchPulsar} we show the first discovered fast radio burst (FRB; \cite{doi:10.1126/science.1147532}) after converting the 13 PSRFITS data files corresponding to the 13 beams of the multi-beam receiver into a single SDHDF file. We note that the data volume in SDHDF is slightly smaller than the original PSRFITS files as we do not require the extra metadata information every 4096 samples.

\section{DISCUSSION}
\label{sec:sec5}
A commonly-held criticism and limitation of HDF is the lack of accessibility and readability --- HDF is an entirely binary data format and not human-readable. Astronomers have historically enjoyed the benefits of partially text-based formats like FITS. For instance, the metadata in those files can be  inspected in a standard text editor. However, we believe that the ability to manually edit a data file goes against good provenance practices, and in the Big Data era it is intractable to open an entire file in a simple text editor. We therefore believe that HDF is a suitable format for storing radio astronomy data for the following reasons:
\begin{itemize}
\item\textbf{Overcoming the limitations of FITS} --- Existing FITS derivatives have limitations, for example RPFITS\footnote{\url{https://www.atnf.csiro.au/computing/software/rpfits.html}} did not completely adhere to the FITS standard and lacked comprehensive metadata, SDFITS suffered from structure variation across files from different observations, and was not suitable for storing or processing of high-volume wide-frequency band data. PSRFITS, although the primary data format for pulsar astronomy data, is not flexible enough for the storage of data and metadata from spectral-line or continuum observing modes. Unlike HDF, the FITS metadata also suffer from keyword truncation, the Data Units (`extensions') do not easily accommodate appending of large data, and there is no support for hierarchical grouping of data and metadata required by multi-beam observations.
\item\textbf{No suitable alternative formats} --- We found no other file formats that could meet our requirements. We investigated Advanced Scientific Data Format (ASDF; \cite{2015A&C....12..240G}), a format designed to find a middle ground between FITS and hierarchically structured content, however whilst overcoming some of the limitations of FITS, there were restrictions on data and metadata storage and it was deemed too generic for storing high-volume radio astronomy data from wide-bandwidths and multiple beams.
\item\textbf{Improved I/O performance and compression} --- The HDF libraries support `chunked' over `contiguous' storage to improve I/O performance. SDHDF files can be cloud-optimised for distributed computing by writing out with increased chunk sizes.
With regards to compression, historically metadata were stored sparsely, as storing multi-dimensional arrays of information such as flags would dramatically increase file size. However, the ability to compress these metadata arrays natively in HDF encourages their inclusion in the file.   \cite{2015A&C....12..212P} demonstrated this improved I/O and compression of HDF over FITS during bench-marking of a FITS to HDF converter. 
\item\textbf{Accessibility} --- Unlike FITS, HDF supports Unicode characters and is therefore suitable for situations where text in languages other than English is required. For instance, we have successfully demonstrated that observer names can be stored in HDF format in Chinese characters.
\end{itemize}

\subsection{Extending the SDHDF format}\label{sec:extension}
The spectral-line and continuum astronomy communities who use Parkes for their observations are adopting the new file format and processing tools. Some teams use their own processing tools, others are now making use of \textsc{INSPECTA}. Developers of external tools are also adopting SDHDF. In particular  AOFlagger\footnote{\url{https://aoflagger.readthedocs.io/en/latest/}}, an RFI flagging package, can now parse and flag SDHDF files.

The SDHDF definition is managed in an open source repository and we encourage feedback on its development. We are currently developing SDHDF version 4.0\footnote{We note that for data written in SDHDF $<$4.0 format, users who wish to work with \textsc{INSPECTA} will need to checkout the `sdhdfv3' branch of the software.}, for which we have made improvements to the naming scheme, dataset dimensions and attributes based on user feedback, while the underlying structure remains unchanged --- important changes that apply to archival data prior to version 4.0 are shown in Tables \ref{tb:sdhdf_updates} and \ref{tb:sdhdf_dim_updates} respectively.

\begin{table}
  \caption{SDHDF naming scheme changes}
  \begin{center}
  \begin{tabular}{llll}
    SDHDF $<$3.0 & SDHDF 3.0 (Released: April 2023) & SDHDF 4.0 (Under development) & HDF type     \\
\hline
\\
config             & config      & configuration            & Group    \\
cal\_data\_on      & cal\_on     & calibrator\_data\_on     & Dataset  \\
cal\_data\_off     & cal\_off    & calibrator\_data\_off    & Dataset  \\
cal32\_data        & cal\_binned & calibrator\_data\_binned & Dataset  \\
cal\_frequency     & frequency   & frequency                & Dataset  \\ 
\hline
data\_weights      & weights     & weights                  & Dataset  \\ 
data\_flags        & flags       & flags                    & Dataset  \\ 
\hline
obs\_params        & obs\_params & observation\_parameters  & Dataset  \\
cal\_obs\_params   & cal\_obs\_params & calibrator\_observation\_parameters  & Dataset \\
\hline   
band\_params       & band\_params & band\_parameters        & Dataset  \\
cal\_band\_params  & cal\_band\_params & calibrator\_band\_parameters        & Dataset \\
\hline
beam\_params       & beam\_params & beam\_parameters        & Dataset  \\
\hline
backend\_config    & backend\_config & intrument\_configuration & Dataset \\
cal\_backend\_config  & cal\_backend\_config & intrument\_configuration & Dataset \\
\\
  \end{tabular}
  \end{center}
  \label{tb:sdhdf_updates}
\end{table}

\begin{table}
  \caption{SDHDF dataset dimension changes}
  \begin{center}
  \begin{tabular}{p{1in}p{1in}p{1in}p{2in}}
    HDF dataset  & SDHDF $<$3.0 & SDHDF $>=$3.0 & Reason \\
\hline
\\
frequency    & 1-dimensional  & 2-dimensional & Allow storage of time-dependent frequency information \\
flags        & 2-dimensional  & 4-dimensional & To match the dimensions of the data, allowing flags to be applied to any dimension of the data\\
weights      & 2-dimensional  & 4-dimensional & To match the dimensions of the data, allowing weights to be applied to any dimension of the data \\
\hline
  \end{tabular}
  \end{center}
  \label{tb:sdhdf_dim_updates}
\end{table}

We do expect further updates as new observing systems are commissioned and more observatories make use of this format. New datasets and attributes will be added in the future. For example, the ability to store statistics from digitisers and a flag for data quality would both be useful additions to the definition and planned for future versions. We note that SDHDF files do not require that all parameters are included. For instance, the observation metadata includes parameters such as the wind speed, which may not be relevant (or even known) for some observations.  Similarly, there will always be cases in which a particular observing system requires metadata that is unique to that system.  The SDHDF format does not preclude the addition of observatory-specific information.

\subsection{Virtual Observatory Support}\label{sec:access}
For the benefit of all users of astronomical data, we would like to see SDHDF become accessible and interoperable with the Virtual Observatory (VO). This can be achieved by following protocols of The International Virtual Observatory Alliance\footnote{\url{https://ivoa.net/}} (IVOA), an organisation that ``debates and agrees the technical standards that are needed to make the VO possible''. The ATOA already has VO interoperability in place for RPFITS format data --- we believe that the variety and extensive metadata included in the SDHDF format provides a significant improvement on this.

\section{CONCLUSION}
\label{sec:concl}
In this paper we have presented a new data format for radio astronomy data; the Spectral-Domain Hierarchical Data Format (SDHDF). The format is flexible enough to store high-time and high frequency-resolution data and metadata products from the era of large telescopes with single pixel or multi-beam receivers.

Murriyang, the CSIRO Parkes 64-metre radio telescope, is the first instrument to use the format natively for all spectral-line and continuum observations. The Ultra-Wide-Bandwidth Low Frequency receiver, a single pixel feed, has produced SDHDF files at Parkes since 2018,  and the cryogenically-cooled phased array feed, to be commissioned in 2024, is expected to use this format for storing the data products from multiple beams.

We have developed the \textsc{INSPECTA} software package for the interrogation, processing and calibration of SDHDF data, and have demonstrated some of its features when working with radio astronomy data.

We envisage other institutions adopting SDHDF as their primary format because of the flexibility to store multiple data streams and comprehensive metadata, and its adherence to a formal definition suitable for long-term archiving.

\section*{Acknowledgements}
The Parkes radio telescope is part of the Australia Telescope National Facility (grid.421683.a) which is funded by the Australian Government for operation as a National Facility managed by CSIRO.
We acknowledge the Wiradjuri people as the Traditional Owners of the Observatory site.
This paper includes archived data obtained through the Australia Telescope Online Archive (http://atoa.atnf.csiro.au).
T.V.W. is supported by an NSF Astronomy and Astrophysics Postdoctoral Fellowship under award AST-2202340.
We thank Dr. Philip Edwards for his invaluable comments on this paper.

\bibliographystyle{elsarticle-num}
\bibliography{refs.bib}
\clearpage

\onecolumn
\appendix
\section{THE SDHDF DEFINITION}
\label{appendix:appdxA}
  \tiny
\begin{longtable}{p{2in}p{0.7in}p{1.5in}p{0.5in}p{1in}}
\textbf{SDHDF Definition Overview} & & & & \\
\hline
SDHDF Definition Version & 4.0 & & \\
Author & Lawrence Toomey & & & \\
Copyright & CSIRO, 2024 & & & \\
\hline
\\ 
\\
\\
\\
\textbf{SDHDF File Overview} & & & & \\
HDF\_Object\_Name & HDF\_Object\_Type & Description & Unit & Value \\
\hline
SDHDF\_CLASS & Attribute & SDHDF class name & None & sdhdf\_file \\
SDHDF\_DESCRIPTION & Attribute & SDHDF object description & None & string \\
\hline
/beam\_N & Group & - & - & - \\
SDHDF\_CLASS & Attribute & SDHDF class name & None & sdhdf\_beam \\
SDHDF\_DESCRIPTION & Attribute & SDHDF object description & None & string \\
\hline
/beam\_N/band\_0 & Group & - & - & - \\
SDHDF\_CLASS & Attribute & SDHDF class name & None & sdhdf\_band \\
SDHDF\_DESCRIPTION & Attribute & SDHDF object description & None & string \\
\hline
/beam\_N/band\_N/astronomy\_data & Group & - & - & - \\
SDHDF\_CLASS & Attribute & SDHDF class name & None & sdhdf\_data \\
SDHDF\_DESCRIPTION & Attribute & SDHDF object description & None & string \\
\hline
/beam\_N/band\_N/astronomy\_data/data & Dataset & - & - & - \\
DATA\_DESCRIPTION & Attribute & Data description (Antenna Temperature, Brightness Temperature, Equivalent Point Source Flux Density, other & None & None \\
DATA\_TYPE & Attribute & Data type & None & float32 \\
DIMENSION\_LABELS & Attribute & HDF dimension labels & None & string \\
FREQUENCY & Attribute & Frequency, data array dimension 2 & None & frequency \\
NORMALISATION\_FACTOR & Attribute & The factor by which the number of samples integrated into each output sample are normalised & None & 0.0 \\
NUMBER\_OF\_BINS & Attribute & Number of phase bins & None & 1 \\
PHASE\_BIN & Attribute & Phase bins, data array dimension 3 & None & bin \\
PRODUCT\_TYPE & Attribute & Product type, data array dimension 1 & None & None \\
SDHDF\_CLASS & Attribute & SDHDF class name & None & sdhdf\_waterfall \\
SDHDF\_DESCRIPTION & Attribute & SDHDF object description & None & string \\
TIME & Attribute & Integrations, data array dimension 0 & None & time \\
UNIT & Attribute & Data unit, the number of native time samples that are integrated into each output time sample & None & counts \\
\hline
/beam\_N/band\_N/astronomy\_data/flags & Dataset & - & - & - \\
DATA\_TYPE & Attribute & Data type & None & boolean \\
DIMENSION\_LABELS & Attribute & HDF dimension labels & None & string \\
FLAG\_TYPE & Attribute & Type of flagging method applied & None & Default: 0=not flagged, 1=flagged) \\
FREQUENCY & Attribute & Frequency, data array dimension 2 & None & frequency \\
PHASE\_BIN & Attribute & Phase bins, data array dimension 3 & None & bin \\
PRODUCT\_TYPE & Attribute & Product type, data array dimension 1 & None & None \\
SDHDF\_CLASS & Attribute & SDHDF class name & None & sdhdf\_flags \\
SDHDF\_DESCRIPTION & Attribute & SDHDF object description & None & Flags dataset \\
TIME & Attribute & Integrations, data array dimension 0 & None & time \\
\hline
/beam\_N/band\_N/astronomy\_data/frequency & Dataset & - & - & - \\
DATA\_TYPE & Attribute & Data type & None & float32 \\
DIMENSION\_LABELS & Attribute & HDF dimension labels & None & string \\
FRAME & Attribute & Spectral reference frame (barycentric, topocentric, LSR, other) & None & topocentric \\
FREQUENCY & Attribute & Frequency, data array dimension 1 & None & frequency \\
SDHDF\_CLASS & Attribute & SDHDF class name & None & sdhdf\_frequency \\
SDHDF\_DESCRIPTION & Attribute & SDHDF object description & None & Channel centre frequencies \\
TIME & Attribute & Integrations, data array dimension 0 & None & time \\
UNIT & Attribute & Data unit & None & MHz \\
\hline
/beam\_N/band\_N/astronomy\_data/weights & Dataset & - & - & - \\
DATA\_TYPE & Attribute & Data type & None & float32 \\
DIMENSION\_LABELS & Attribute & HDF dimension labels & None & string \\
FREQUENCY & Attribute & Frequency, data array dimension 2 & None & frequency \\
PHASE\_BIN & Attribute & Phase bins, data array dimension 3 & None & bin \\
PRODUCT\_TYPE & Attribute & Product type, data array dimension 1 & None & None \\
SDHDF\_CLASS & Attribute & SDHDF class name & None & sdhdf\_weights \\
SDHDF\_DESCRIPTION & Attribute & SDHDF object description & None & Weights dataset \\
TIME & Attribute & Integrations, data array dimension 0 & None & time \\
UNIT & Attribute & Data unit & None & dimensionless \\
WEIGHT\_TYPE & Attribute & Type of weighting method applied & None & Product of the channel bandwidth (Hz) and integration time (s) \\
\hline
/beam\_N/band\_N/calibrator\_data & Group & - & - & - \\
SDHDF\_CLASS & Attribute & SDHDF class name & None & sdhdf\_data \\
SDHDF\_DESCRIPTION & Attribute & SDHDF object description & None & string \\
\hline
/beam\_N/band\_N/calibrator\_data/data & Dataset & - & - & - \\
DATA\_DESCRIPTION & Attribute & Data description (Antenna Temperature, Brightness Temperature, Equivalent Point Source Flux Density, other & None & None \\
DATA\_TYPE & Attribute & Data type & None & float32 \\
DIMENSION\_LABELS & Attribute & HDF dimension labels & None & string \\
FREQUENCY & Attribute & Frequency, data array dimension 2 & None & frequency \\
NORMALISATION\_FACTOR & Attribute & The factor by which the number of samples integrated into each output bin are normalised & None & 0.0 \\
NUMBER\_OF\_BINS & Attribute & Number of phase bins & None & 1 \\
NUMBER\_OF\_CHANNELS & Attribute & Number of frequency channels & None & 0 \\
PHASE\_BIN & Attribute & Phase bins, data array dimension 3 & None & bin \\
PRODUCT\_TYPE & Attribute & Product type, data array dimension 1 & None & None \\
SDHDF\_CLASS & Attribute & SDHDF class name & None & sdhdf\_waterfall \\
SDHDF\_DESCRIPTION & Attribute & SDHDF object description & None & string \\
TIME & Attribute & Integrations, data array dimension 0 & None & time \\
UNIT & Attribute & Data unit, the number of native time samples that areintegrated into each output time sample & None & counts \\
\hline
/beam\_N/band\_N/calibrator\_data/flags & Dataset & - & - & - \\
DATA\_TYPE & Attribute & Data type & None & boolean \\
DIMENSION\_LABELS & Attribute & HDF dimension labels & None & string \\
FLAG\_TYPE & Attribute & Type of flagging method applied & None & Default: 0=not flagged, 1=flagged) \\
FREQUENCY & Attribute & Frequency, data array dimension 2 & None & frequency \\
PHASE\_BIN & Attribute & Phase bins, data array dimension 3 & None & bin \\
PRODUCT\_TYPE & Attribute & Product type, data array dimension 1 & None & None \\
SDHDF\_CLASS & Attribute & SDHDF class name & None & sdhdf\_flags \\
SDHDF\_DESCRIPTION & Attribute & SDHDF object description & None & Flags dataset \\
TIME & Attribute & Integrations, data array dimension 0 & None & time \\
\hline
/beam\_N/band\_N/calibrator\_data/frequency & Dataset & - & - & - \\
DATA\_TYPE & Attribute & Data type & None & float32 \\
DIMENSION\_LABELS & Attribute & HDF dimension labels & None & string \\
FRAME & Attribute & Spectral reference frame (barycentric, topocentric, LSR, other) & None & topocentric \\
FREQUENCY & Attribute & Frequency, data array dimension 1 & None & frequency \\
SDHDF\_CLASS & Attribute & SDHDF class name & None & sdhdf\_frequency \\
SDHDF\_DESCRIPTION & Attribute & SDHDF object description & None & Channel centre frequencies \\
TIME & Attribute & Integrations, data array dimension 0 & None & time \\
UNIT & Attribute & Data unit & None & MHz \\
\hline
/beam\_N/band\_N/calibrator\_data/weights & Dataset & - & - & - \\
DATA\_TYPE & Attribute & Data type & None & float32 \\
DIMENSION\_LABELS & Attribute & HDF dimension labels & None & string \\
FREQUENCY & Attribute & Frequency, data array dimension 2 & None & frequency \\
PHASE\_BIN & Attribute & Phase bins, data array dimension 3 & None & bin \\
PRODUCT\_TYPE & Attribute & Product type, data array dimension 1 & None & None \\
SDHDF\_CLASS & Attribute & SDHDF class name & None & sdhdf\_weights \\
SDHDF\_DESCRIPTION & Attribute & SDHDF object description & None & Weights dataset \\
TIME & Attribute & Integrations, data array dimension 0 & None & time \\
UNIT & Attribute & Data unit & None & dimensionless \\
WEIGHT\_TYPE & Attribute & Type of weighting method applied & None & Product of the channel bandwidth (Hz) and integration time (s) \\
\hline
/beam\_N/band\_N/metadata & Group & - & - & - \\
SDHDF\_CLASS & Attribute & SDHDF class name & None & sdhdf\_metadata \\
SDHDF\_DESCRIPTION & Attribute & SDHDF object description & None & string \\
\hline
/beam\_N/band\_N/metadata/calibrator\_observation
\_parameters & Dataset & - & - & - \\
SDHDF\_CLASS & Attribute & SDHDF class name & None & sdhdf\_table \\
SDHDF\_DESCRIPTION & Attribute & SDHDF object description & None & Metadata specific to the observation integrations \\
\hline
/beam\_N/band\_N/metadata/observation
\_parameters & Dataset & - & - & - \\
SDHDF\_CLASS & Attribute & SDHDF class name & None & sdhdf\_table \\
SDHDF\_DESCRIPTION & Attribute & SDHDF object description & None & Metadata specific to the observation integrations \\
\hline
/beam\_N/metadata & Group & - & - & - \\
SDHDF\_CLASS & Attribute & SDHDF class name & None & sdhdf\_metadata \\
SDHDF\_DESCRIPTION & Attribute & SDHDF object description & None & string \\
\hline
/beam\_N/metadata/band\_parameters & Dataset & - & - & - \\
CENTRE\_FREQUENCY & Attribute & Band centre frequency & MHz & 0.0 \\
HIGH\_FREQUENCY & Attribute & Band range high frequency & MHz & 0.0 \\
LABEL & Attribute & Band label & None & None \\
LOW\_FREQUENCY & Attribute & Band range low frequency & MHz & 0.0 \\
NUMBER\_OF\_BINS & Attribute & Number of phase bins & None & 1 \\
NUMBER\_OF\_CHANNELS & Attribute & Number of channels in band & None & 0 \\
NUMBER\_OF\_INTEGRATIONS & Attribute & Number of spectral data integrations in band & None & 0 \\
NUMBER\_OF\_POLARISATIONS & Attribute & Number of polarisations (1, 2, 4) & None & 0 \\
PARTIAL\_NUMBER\_OF\_INTEGRATIONS & Attribute & Number of partial or incomplete spectral data integrations in band & None & 0 \\
POLARISATION\_TYPE & Attribute & Polarisation type (I, Q, U, V, RR, LL RL LR, XX, YY, XY, YX, POLI, POLA, AABBCRCI (for 4 pol coherence data, where AA and BB are the direct products of the two input channels A and B, and CR and CI are the real and imaginary parts of the cross product A* B), AA+BB (for 1 pol data with summed orthogonal products), AABB (for 2 pol data) & None & None \\
REQUESTED\_INTEGRATION\_TIME & Attribute & Requested integration time & s & 0.0 \\
SDHDF\_CLASS & Attribute & SDHDF class name & None & sdhdf\_table \\
SDHDF\_DESCRIPTION & Attribute & SDHDF object description & None & Metadata specific to the frequency bands of the antenna beam \\
\hline
/beam\_N/metadata/calibrator\_band\_parameters & Dataset & - & - & - \\
CENTRE\_FREQUENCY & Attribute & Band centre frequency & MHz & 0.0 \\
HIGH\_FREQUENCY & Attribute & Band range high frequency & MHz & 0.0 \\
LABEL & Attribute & Band label & None & None \\
LOW\_FREQUENCY & Attribute & Band range low frequency & MHz & 0.0 \\
NUMBER\_OF\_BINS & Attribute & Number of phase bins & None & 1 \\
NUMBER\_OF\_CHANNELS & Attribute & Number of channels in band & None & 0 \\
NUMBER\_OF\_INTEGRATIONS & Attribute & Number of spectral data integrations in band & None & 0 \\
NUMBER\_OF\_POLARISATIONS & Attribute & Number of polarisations (1, 2, 4) & None & 0 \\
PARTIAL\_NUMBER\_OF\_INTEGRATIONS & Attribute & Number of partial or incomplete spectral data integrations in band & None & 0 \\
POLARISATION\_TYPE & Attribute & Polarisation type (I, Q, U, V, RR, LL RL LR, XX, YY, XY, YX, POLI, POLA, AABBCRCI (for 4 pol coherence data, where AA and BB are the direct products of the two input channels A and B, and CR and CI are the real and imaginary parts of the cross product A* B), AA+BB (for 1 pol data with summed orthogonal products), AABB (for 2 pol data) & None & None \\
REQUESTED\_INTEGRATION\_TIME & Attribute & Requested integration time & s & 0.0 \\
SDHDF\_CLASS & Attribute & SDHDF class name & None & sdhdf\_table \\
SDHDF\_DESCRIPTION & Attribute & SDHDF object description & None & Metadata specific to the frequency bands of the antenna beam \\
\hline
/configuration & Group & - & - & - \\
SDHDF\_CLASS & Attribute & SDHDF class name & None & sdhdf\_configuration \\
SDHDF\_DESCRIPTION & Attribute & SDHDF object description & None & string \\
\hline
/configuration/instrument\_configuration & Dataset & - & - & - \\
SDHDF\_CLASS & Attribute & SDHDF class name & None & sdhdf\_table \\
SDHDF\_DESCRIPTION & Attribute & SDHDF object description & None & Digital backend instrument configuration \\
\hline
/configuration/receiver\_configuration & Dataset & - & - & - \\
SDHDF\_CLASS & Attribute & SDHDF class name & None & sdhdf\_table \\
SDHDF\_DESCRIPTION & Attribute & SDHDF object description & None & Receiver instrument configuration \\
\hline
/configuration/telescope\_configuration & Dataset & - & - & - \\
SDHDF\_CLASS & Attribute & SDHDF class name & None & sdhdf\_table \\
SDHDF\_DESCRIPTION & Attribute & SDHDF object description & None & Telescope configuration \\
\hline
/metadata & Group & - & - & - \\
SDHDF\_CLASS & Attribute & SDHDF class name & None & sdhdf\_metadata \\
SDHDF\_DESCRIPTION & Attribute & SDHDF object description & None & string \\
\hline
/metadata/beam\_parameters & Dataset & - & - & - \\
DECLINATION & Attribute & Coordinates in Declination & DD:MM:SS.s & None \\
EQUINOX & Attribute & Equinox of the coordinates & Julian year & 2000 \\
LABEL & Attribute & Beam label & None & None \\
NUMBER\_OF\_BANDS & Attribute & Number of frequency bands & None & 0 \\
RIGHT\_ASCENSION & Attribute & Coordinates in Right Ascension & HH:MM:SS.s & None \\
SDHDF\_CLASS & Attribute & SDHDF class name & None & sdhdf\_table \\
SDHDF\_DESCRIPTION & Attribute & SDHDF object description & None & Metadata specific to the antenna beam \\
SOURCE & Attribute & Source name & None & None \\
\hline
/metadata/history & Dataset & - & - & - \\
DATE & Attribute & Date (UTC YYYY-MM-DD-hh:mm:ss) & None & None \\
PROCESS & Attribute & Process name & None & None \\
PROCESSING\_HOST & Attribute & Host machine running the process & None & None \\
PROCESS\_ARGUMENTS & Attribute & Process command arguments & None & None \\
PROCESS\_DESCRIPTION & Attribute & Process description & None & None \\
PROCESS\_LOG & Attribute & Logged output from the process & None & None \\
SDHDF\_CLASS & Attribute & SDHDF class name & None & sdhdf\_table \\
SDHDF\_DESCRIPTION & Attribute & SDHDF object description & None & Metadata specific to the processing history of the file \\
\hline
/metadata/primary\_header & Dataset & - & - & - \\
CALIBRATION\_MODE & Attribute & Calibration mode (ON: calibrator data recorded, OFF: no calibrator data recorded) & None & None \\
DATE & Attribute & File creation date (UTC YYYY-MM-DD-hh:mm:ss) & YYYY-MM-DD-hh:mm:ss & None \\
FILE\_FORMAT & Attribute & File format & None & HDF \\
FILE\_FORMAT\_VERSION & Attribute & File format version & None & 5.0 \\
HEADER\_DEFINITION & Attribute & File format definition & None & string \\
HEADER\_DEFINITION\_VERSION & Attribute & File format definition version & None & 4.0 \\
INSTRUMENT & Attribute & Backend instrument name & None & None \\
NUMBER\_OF\_BEAMS & Attribute & Number of beams & None & 0 \\
OBSERVATION\_TYPE & Attribute & Observation type (TRACK, SCAN) & None & None \\
OBSERVER & Attribute & Observer name & None & None \\
PROJECT\_ID & Attribute & Project identifier & None & None \\
RECEIVER & Attribute & Receiver name & None & None \\
SCHEDULE\_ID & Attribute & Observing schedule identifier & None & 0 \\
SDHDF\_CLASS & Attribute & SDHDF class name & None & sdhdf\_table \\
SDHDF\_DESCRIPTION & Attribute & SDHDF object description & None & General observation metadata \\
TELESCOPE & Attribute & Telescope name & None & None \\
UTC\_START & Attribute & Observation start (UTC YYYY-MM-DD-hh:mm:ss) & YYYY-MM-DD-hh:mm:ss & None \\
\hline
/metadata/schedule & Dataset & - & - & - \\
SDHDF\_CLASS & Attribute & SDHDF class name & None & sdhdf\_table \\
SDHDF\_DESCRIPTION & Attribute & SDHDF object description & None & Metadata specific to the scheduling of the observation \\
\hline
/metadata/software\_versions & Dataset & - & - & - \\
PROCESS & Attribute & Process name & None & None \\
SDHDF\_CLASS & Attribute & SDHDF class name & None & sdhdf\_table \\
SDHDF\_DESCRIPTION & Attribute & SDHDF object description & None & Metadata specific to software packages used for creating or processing the file \\
SOFTWARE & Attribute & Software package & None & None \\
SOFTWARE\_DESCRIPTION & Attribute & Software package description & None & None \\
SOFTWARE\_VERSION & Attribute & Software package version & None & None \\
\hline
\end{longtable}

\clearpage

\section{Example Observation Parameters}
\label{appendix:appdxB}
  \tiny
\begin{longtable}{p{2.1in}p{0.7in}p{1.6in}p{0.5in}p{0.5in}}
HDF\_Object\_Name & HDF\_Object\_Type & Description & Unit & Data\_Type \\
\hline
/beam\_N/band\_N/observation\_parameters & Dataset & - & - & - \\
EQUINOX & Attribute & Equinox of the coordinates & Julian year & string \\
SDHDF\_CLASS & Attribute & SDHDF class name & None & string \\
SDHDF\_DESCRIPTION & Attribute & SDHDF object description & None & string \\
SOURCE\_LABEL & Attribute & Source label for multiple sources, e.g. for scans & None & string \\
AZIMUTH\_ANGLE & Attribute & Antenna azimuth angle & degrees & float64 \\
DECLINATION & Attribute & Declination & DD:MM:SS.s & string \\
ELEVATION\_ANGLE & Attribute & Antenna elevation angle & degrees & float64 \\
ELAPSED\_TIME & Attribute & Elapsed time from UTC\_START, at integration centre & s & float64 \\
DRIVE\_STATUS & Attribute & Telescope drive status (STATIONARY = SLEWING = STOPPING = PARKING = PARKED = STOWED = 0, TRACKING = 1, RAMPING UP = 2 RAMPING DOWN = 2, SCANNING = 3) & NULL & int64 \\
GALACTIC\_LATITUDE & Attribute & Galactic latitude & degrees & float64 \\
GALACTIC\_LONGITUDE & Attribute & Galactic longitude & degrees & float64 \\
HOUR\_ANGLE & Attribute & Hour angle & degrees & float64 \\
INTEGRATION\_TIME & Attribute & Actual integration time & s & float64 \\
LOCAL\_TIME & Attribute & Local time at the observatory & HH:MM:SS.s & string \\
MJD & Attribute & Timestamp at integration centre (Modified Julian Date) & days & int64 \\
FRACTIONAL\_MJD & Attribute & Time at integration centre since start of the MJD & seconds & float64 \\
PARALLACTIC\_ANGLE & Attribute & Parallactic angle & degrees & float64 \\
PRESSURE & Attribute & Atmospheric pressure & hPa & float64 \\
PRESSURE\_MSL & Attribute & Atmospheric pressure at mean sea level & hPa & float64 \\
RIGHT\_ASCENSION & Attribute & Right ascension & HH:MM:SS.s & string \\
RELATIVE\_HUMIDITY & Attribute & Outside relative humidity & \% & float64 \\
TEMPERATURE & Attribute & Outside temperature & degrees C & float64 \\
UTC & Attribute & Timestamp at integration centre (UTC) & HH:MM:SS.s & string \\
WIND\_DIRECTION & Attribute & Wind direction & degrees & float64 \\
WIND\_SPEED & Attribute & Wind speed & km/hr & float64 \\
ZENITH\_ANGLE & Attribute & Antenna zenith angle & degrees & float64 \\
\hline
\end{longtable}
 
\end{document}